\documentclass{article}
\usepackage{spconf,amsmath,graphicx,subcaption,multirow,booktabs,hyperref}

\newcommand\blfootnote[1]{% credit: https://tex.stackexchange.com/questions/30720/footnote-without-a-marker
  \begingroup
  \renewcommand\thefootnote{}\footnote{#1}%
  \addtocounter{footnote}{-1}%
  \endgroup
}
\title{Two-Step Color-Polarization Demosaicking Network}
\name{Vy Nguyen, Masayuki Tanaka, Yusuke Monno, and Masatoshi Okutomi}
\address{Tokyo Institute of Technology, Tokyo, Japan}

\begin{document}
\maketitle
\begin{abstract}
Polarization information of light in a scene is valuable for various image processing and computer vision tasks. A division-of-focal-plane polarimeter is a promising approach to capture the polarization images of different orientations in one shot, while it requires color-polarization demosaicking. In this paper, we propose a two-step color-polarization demosaicking network~(TCPDNet), which consists of two sub-tasks of color demosaicking and polarization demosaicking. We also introduce a reconstruction loss in the YCbCr color space to improve the performance of TCPDNet. Experimental comparisons demonstrate that TCPDNet outperforms existing methods in terms of the image quality of polarization images and the accuracy of Stokes parameters.
\end{abstract}
\blfootnote{© 2022 IEEE. Personal use of this material is permitted. Permission from IEEE must be obtained for all other uses, in any current or future media, including reprinting/republishing this material for advertising or promotional purposes, creating new collective works, for resale or redistribution to servers or lists, or reuse of any copyrighted component of this work in other works.}
\section{Introduction}
\label{sec:intro}

Polarization information of light in a scene can be helpful in various image processing and computer vision tasks such as transparent object segmentation~\cite{kalra2020deep}, 3D reconstruction~\cite{zhao2020polarimetric}, and specular removal~\cite{jospin2018embedded}. A polarization image can be obtained by placing a polarizer in front of the camera lens, where the images of different polarization orientations of the same scene can be obtained by rotating the polarizer. Four polarization images of $0^{\circ}$, $45^{\circ}$, $90^{\circ}$, and $135^{\circ}$ are typically captured to robustly infer necessary polarization information, e.g., Stokes parameters, the angle of polarization~(AoP), and the degree of polarization~(DoP)~\cite{huynh2013shape}.

There are two popular polarization image acquisition approaches~\cite{tyo2006review}: a division-of-time polarimeter and a  division-of-focal-plane polarimeter. In the division-of-time polarimeter, a linear polarizer in front of the camera lens is sequentially rotated in time to obtain different polarization states of a pixel. However, the division-of-time polarimeter is only applicable to static scenes where no camera and object movement exists. In the division-of-focal-plane polarimeter, a micro-polarizer array is used to capture different polarization orientations as a mosaic pattern at once. Each pixel of the captured mosaic pattern only contains the information of one polarization orientation. The division-of-focal-plane polarimeter can be applied to dynamic scenes but it requires a demosaicking process, which is an interpolation process of missing pixel values. The demosaicking process strongly affects the image quality.

\begin{figure}
     \centering
     \begin{subfigure}[b]{0.48\textwidth}
         \centering
         \includegraphics[width=\textwidth]{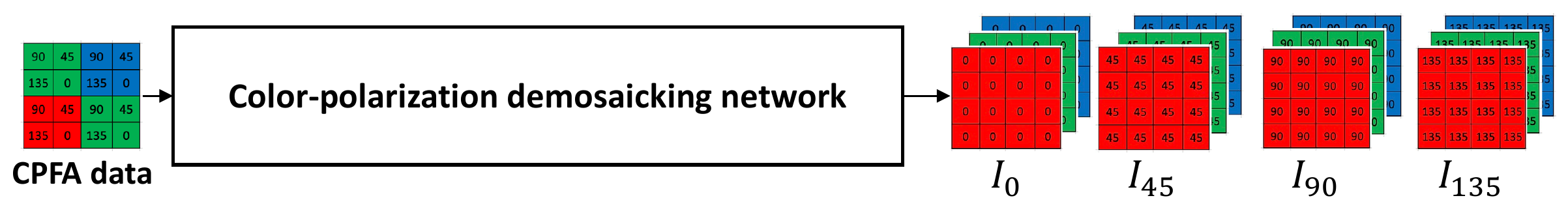}
         \vspace*{-1.5em} 
         \caption{Existing single-step approach}
         \vspace*{1em}
     \end{subfigure}
     \vfill
     \begin{subfigure}[b]{0.48\textwidth}
         \centering
         \includegraphics[width=\textwidth]{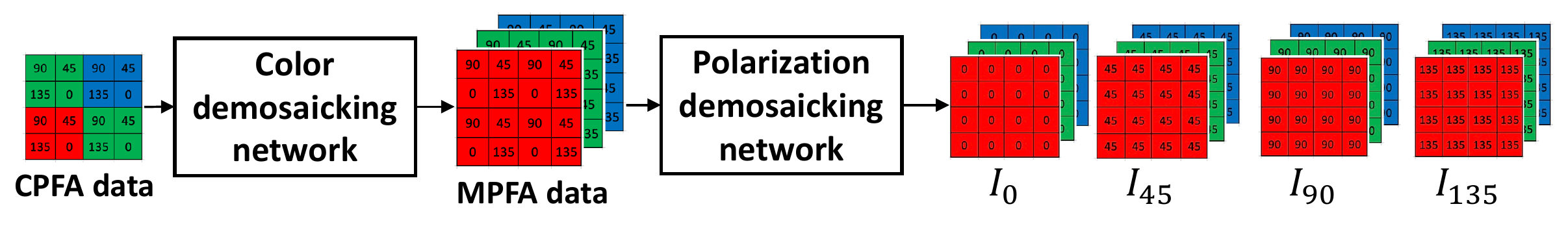}
         \vspace*{-1.5em}
         \caption{Our two-step approach}
     \end{subfigure}
        \caption{Two network-based approaches for color-polarization demosaicking: (a) Existing single-step approach and (b) our two-step approach, which consists of color demosaicking and polarization demosaicking.}
        \label{fig:relatedwork}
\end{figure}

\begin{figure*}
\centerline{\includegraphics[width=\linewidth]{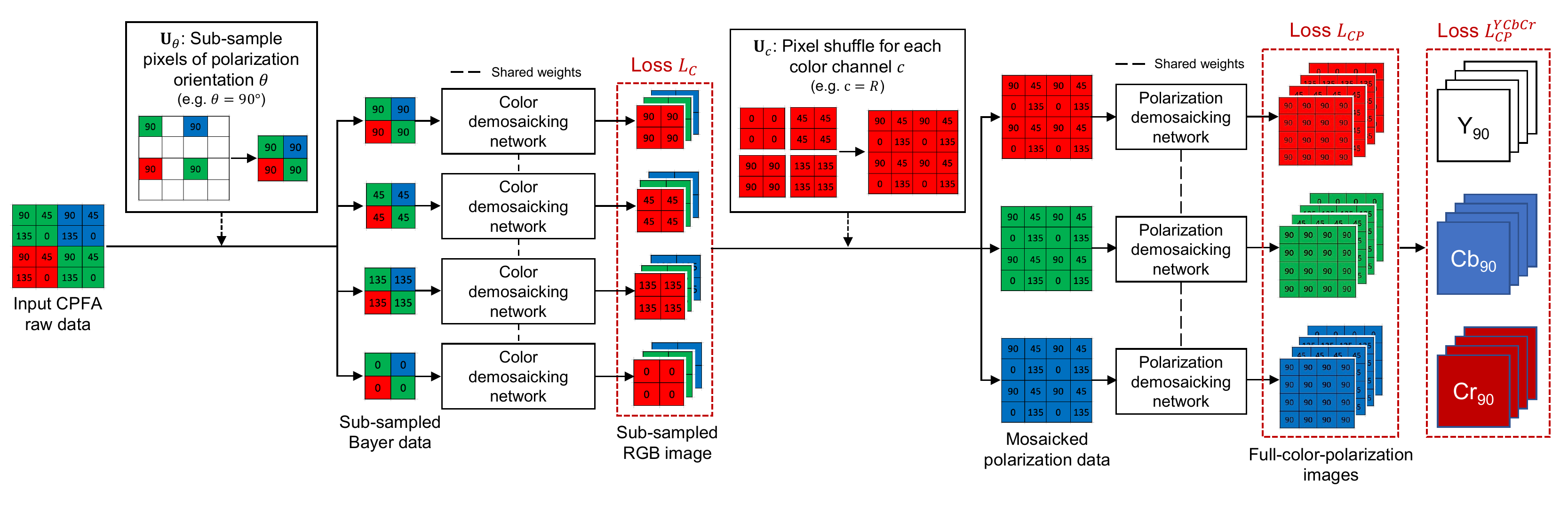}}
\vspace{-3mm}
\caption{The overall pipeline of our Two-step Color-Polarization Demosaicking Network (TCPDNet).}
\label{fig:2step}
\end{figure*}

In this paper, we focus on color-polarization filter array (CPFA) demosaicking for a color division-of-focal-plane polarimeter, such as a Sony sensor~\cite{maruyama20183}. Typically, the input of the CPFA demosaicking is 1-chanel CPFA raw data and the output is 12-channel full-color-polarization images, as shown in Fig.~\ref{fig:relatedwork}. Each pixel of the input 1-channel CPFA raw data contains the intensity of one polarization orientation and one color channel. The output 12-channel full-color-polarization images are a set of RGB and four polarization images with the angles of $0^{\circ}$, $45^{\circ}$, $90^{\circ}$, and $135^{\circ}$. The CPFA demosaicking is more challenging than color filter array (CFA) demosaicking and monochrome polarization filter array (MPFA) demosaicking, because the sampling density of each color-polarization component in the CPFA raw data is very low.

There are two typical approaches of the CPFA demosaicking problem: An interpolation-based approach~\cite{morimatsu2020monochrome, 9583277} and a learning-based approach~\cite{wen2021sparse,wen2019convolutional,sun2021color}. The interpolation-based approach is simple, but the performance is usually limited. Because the learning-based approach with a deep neural network has great potential to obtain high performance, we adopt a network-based approach in this paper.

In the literature, there are two representative learning-based CPFA demosaiking methods: CPDNet~\cite{wen2019convolutional} and CPDCNN~\cite{sun2021color}. CPDNet is an end-to-end network which directly processes the CPFA raw data~\cite{wen2019convolutional}. CPDCNN is a two-branch network which includes the sub-networks to exploit inter-channel correlations and high-frequency information~\cite{sun2021color}. Both of them process the color and the polarization information in an intertwined manner.

In this work, we propose a two-step color-polarization demosaicking network (TCPDNet). The network architecture of TCPDNet is inspired by Morimatsu's interpolation-based CPFA demosaicking method~\cite{morimatsu2020monochrome, 9583277}. Our method consists of two well-studied sub-tasks: color demosaicking~\cite{kiku2016beyond,monno2017adaptive} and polarization demosaicking~\cite{mihoubi2018survey, ahmed2018four, jiang2019minimized}, as shown in Fig.~\ref{fig:relatedwork}(b). In each sub-task, demosaicking network parameters are shared across different kinds of input mosaicked data. We consider that even if the input mosaicked data differ in polarization orientations (for the color demosaicking task) or color channels (for the polarization demosaicking task), their inter-channel correlations should be the same. For example, the color demosaicking networks for $0^{\circ}$ of Bayer CFA mosaicked data and $90^{\circ}$ of Bayer CFA mosaicked data share the same parameters because their RGB correlations should be the same regardless of the polarization orientations. We further improve the performance of TCPDNet by introducing a reconstruction loss in the YCbCr color space. Using the loss function in the YCbCr color space, we expect the demosaicking networks to learn the inter-channel correlations effectively. Experimental results show that our TCPDNet outperforms existing methods by a large margin both quantitatively and qualitatively on Tokyo Tech dataset~\cite{morimatsu2020monochrome, 9583277} and CPDNet dataset~\cite{wen2019convolutional}.

\section{Proposed method}
\label{sec:proposed}

\subsection{Network architecture}
\label{sec:network}

Proposed TCPDNet consists of two sub-networks: color demosaicking network and polarization demosaicking network. This is inspired by the work of Morimatsu et al.~\cite{morimatsu2020monochrome, 9583277}, which proposes a two-step interpolation-based color-polarization demosaicking method. Figure~\ref{fig:2step} shows the overall pipeline of our proposed method. The input CPFA raw data are firstly re-arranged into four sub-sampled Bayer mosaiced data. The sub-sampled Bayer mosaicked data has half size of input raw data in height and in width. The color demosaicking networks then demosaick these four sub-sampled Bayer mosaiced data to generate four sub-sampled RGB images, where the weights of the color demosaicking networks are shared. Then, three mosaicked polarization data are generated by the same manner as the pixel shuffle operation~\cite{shi2016real}. Then, polarization demosaicking networks demosaick those three mosaicked polarization data to generate full-RGB-polarization images which form the output 12-channel data, where the weights of the polarization demosaicking networks are also shared.

Color demosaicking network and polarization demosaicking network have the same demosaicking strategy. Given the input 1-channel mosaicked data, we first interpolate the mosaicked data by bilinear interpolation and then refine the interpolated image by CNN, for which we adopt a high-performance network of U-net~\cite{ronneberger2015u} with the skip connection, though any kind of CNN architectures can be applied. 

\subsection{Loss function}
\label{sec:loss}

We evaluate reconstruction losses with two kinds of images: sub-sampled RGB images and full-color-polarization images, in comparison with the ground-truth data. We refer to the reconstruction losses of the sub-sampled RGB images and the full-color-polarization images as $L_{\text{C}}$ and $L_{\text{CP}}$, respectively, as described in~Fig.~\ref{fig:2step}. 

In this paper, we use the L1-norm for the reconstruction losses. We also introduce $L_{\text{CP}}^{\text{YCbCr}}$, which is the reconstruction loss of the full-color-polarization images in the YCbCr color space, as shown in~Fig.~\ref{fig:2step}.
The backpropagation training with the  $L_{\text{CP}}$ loss in the RGB color space updates the weights of the polarization demosaicking network in a channel-by-channel manner, while the training with the $L_{\text{CP}}^{\text{YCbCr}}$ loss in the YCbCr color space is expected to take account of inter-channel correlations.

Let $\textbf{X}$ be the input CPFA raw data, and $\textbf{Z}$ be the ground-truth 12-channel full-color-polarization data. Let $H$ be the input height and $W$ be the input width. Let ${\bf g}_{\Phi}$ be the color demosaicking network with the parameter $\bf{\Phi}$ and ${\bf h}_{\Psi}$ be the polarization demosaicking network with the parameter $\bf{\Psi}$. Let $\bf{\Theta}$ be $\{0^{\circ}, 45^{\circ}, 90^{\circ}, 135^{\circ}\}$ which is a set of polarization orientations.

For the color demosaicking network, let $\bf{U_{\theta}}$ ($\theta \in \bf{\Theta}$) be the process which sub-samples pixels of polarization orientation $\theta$ from $\textbf{X}$ to form 1-channel sub-sampled Bayer CFA mosaicked data, let $\bf{V_{\theta}}$ be the process which extracts sub-sampled RGB image for polarization angle $\theta$ from the full-color-polarization images $\textbf{Z}$. The reconstruction loss of the sub-sampled RGB images can be expressed as

\begin{equation}L_{\text{C}}({\bf \Phi} ; \textbf{X}, \textbf{Z}) = \frac{1}{3HW}\sum_{\theta \in {\bf \Theta}} \left \| {\bf g}_{\Phi}(\bf{U_{\theta}}(\textbf{X})) - \bf{V_{\theta}}(\textbf{Z}) \right\|_{1}.\end{equation}

For the polarization demosaicking network, let $\textbf{Y}(\textbf{X} ; {\bf \Phi}) = \{ \bf{g_{\rm \Phi}}(U_{\theta}(X)) | \theta \in \Theta \}$ be a set of the outputs of the color demosaicking network. Let ${\bf U}_{c}$ ($c \in \{ R, G, B \}$) be the pixel shuffle operation for $c$-channel of $\textbf{Y}(\textbf{X} ; {\bf \Phi})$. Let $\bf{V_{\rm c}}$ be the process to extract $c$-channel from the full-color-polarization images $\textbf{Z}$. The reconstruction loss of the full-color-polarization images can be expressed as

\begin{equation}\label{eq:lcp}\begin{split}L_{\text{CP}}(&{\bf \Phi}, {\bf \Psi}; \textbf{X}, \textbf{Z}) =\\ &\frac{1}{12HW}\sum_{c \in \{ R, G, B \}}\left \| \bf{h_{\rm \Psi}}(\bf{U_{\rm c}}(\textbf{Y}(\textbf{X} ; \Phi))) - \bf{V_{\rm c}}(\textbf{Z})  \right \|_{1}.\end{split}\end{equation}

The reconstruction loss in Eq.~\ref{eq:lcp} is evaluated in the RGB color space. We then introduce the reconstruction loss in the YCbCr color splace. Let $\bf{A}$ be the function which converts 12-channel full-color-polarization images from the RGB color space to the YCbCr color space and let $Con_c$ be the operation that concatenates each-channel result to form full-color-polarization images. The reconstruction loss in the YCbCr color space can be expressed as

\begin{equation}\begin{split}L_{\text{CP}}^{\text{YCbCr}}(&{\bf \Phi}, {\bf \Psi} ; \textbf{X}, \textbf{Z}) = \\ &\frac{1}{12HW} \left \| \bf{A}({\rm \it Con_c}({\bf h}_{\rm \Psi}({\bf U}_{\rm c}(\textbf{Y}(\textbf{X} ; \Phi))))) - \bf{A}(\textbf{Z}) \right \|_{1}.\end{split}\end{equation}

Then, our proposed loss function can be expressed by combining two reconstruction losses as

\begin{equation}L = L_{\text{C}} + \alpha L_{\text{CP}}^{\text{YCbCr}}
\,,
\end{equation}

where $\alpha$ is a hyperparameter.

In this paper, we experientially set 4 for $\alpha$. The experimental results later show that the loss combination $L_{\text{C}} + 4L_{\text{CP}}^{\text{YCbCr}}$ significantly outperforms the loss combination $L_{\text{C}} + 4L_{\text{CP}}$, which proves the effectiveness of $L_{\text{CP}}^{\text{YCbCr}}$ over $L_{\text{CP}}$.

\section{Experimental Results}
\label{sec:exp}

\begin{table}
\caption{Ablation study on different loss combinations of TCPDNet and the single-step network on Tokyo Tech dataset.}
\label{tab:ablation_morimatsu}
\vspace{-2mm}
\centering
\fontsize{8.5}{11}\selectfont\begin{tabular}{llccc}
\toprule
\multirow{2}{*}{Method} & \multirow{2}{*}{Loss} &  \multicolumn{2}{c}{CPSNR} & Angle error \\ & & $S_0$ & DoP & AoP\\
\midrule
\multirow{2}{*}{Single-step}  & $L_{\text{CP}}$ & 44.72 & 37.03 & 14.30  \\
& $L_{\text{CP}}^{\text{YCbCr}}$ & 44.58 & 36.71 & 14.63 \\
\midrule
\multirow{2}{*}{TCPDNet} & $L_{\text{C}} + 4L_{\text{CP}}$ & 44.59 & 38.74 & 12.70 \\
 & $L_{\text{C}} + 4L_{\text{CP}}^{\text{YCbCr}}$ & \textbf{44.91} & \textbf{38.74} & \textbf{12.65} \\
\bottomrule
\end{tabular}
\vspace{1ex}
\end{table} 

\begin{table*}
\caption{Performance comparison on Tokyo Tech dataset.}
\vspace{-2mm}
\label{tab:sota_morimatsu}
\centering\begin{tabular}{lccccccccc}
\toprule
\multirow{2}{*}{Method} & \multicolumn{8}{c}{CPSNR} & Angle error \\
& $I_0$ & $I_{45}$ & $I_{90}$ & $I_{135}$ & $S_0$ & $S_1$ & $S_2$ & DoP & AoP \\
\midrule
Bilinear interpolation & 34.64 & 34.27 & 35.19 & 34.46 & 36.01 & 42.05 & 39.93 & 30.33 & 23.70 \\
EARI~\cite{morimatsu2020monochrome} & 38.33 & 37.58 & 39.00 & 37.77 & 39.81 & 45.47 & 42.82 & 32.95 & 20.54 \\
IGRI2~\cite{9583277} & 38.40 & 37.59 & 39.07 & 37.78 & 39.60 & 46.38 & 43.05 & 33.17 & 20.05 \\
CPDNet (original)~\cite{wen2019convolutional} & 23.02 & 24.26 & 24.33 & 24.43 & 24.64 & 32.35 & 38.96 & 24.85 & 50.42 \\
CPDNet (re-trained)~\cite{wen2019convolutional} & 28.01 & 27.81 & 28.10 & 27.81 & 28.23 & 45.23 & 41.84 & 31.24 & 32.32 \\
\midrule
TCPDNet (Ours) & \textbf{43.73} & \textbf{43.16} & \textbf{44.46} & \textbf{43.31} & \textbf{44.91} & \textbf{52.82} & \textbf{48.86} & \textbf{38.74} & \textbf{12.65} \\
\bottomrule
\end{tabular}
\vspace{1ex}
\end{table*}

\begin{figure*}
\centerline{\includegraphics[width=\linewidth]{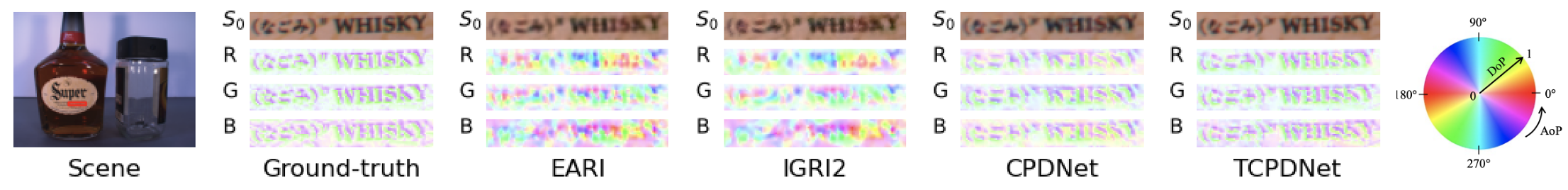}}
\vspace{-2mm}
\caption{Qualitative comparison between our proposed TCPDNet and existing methods. The scene is from Tokyo Tech dataset.}
\label{fig:comparison}
\end{figure*}

\subsection{Datasets and metrics}
\label{sec:dataset}

We evaluated the proposed network with two publicly available datasets: Tokyo Tech dataset~\cite{morimatsu2020monochrome} and CPDNet dataset~\cite{wen2019convolutional}. Due to the page limitation, we only show the results on Tokyo Tech dataset in this paper. Additional results on Tokyo Tech and CPDNet datasets can be seen as a supplemental material in our project page\footnote{\url{http://www.ok.sc.e.titech.ac.jp/res/PolarDem/TCPDNet.html}}. In those two datasets, the ground-truth 12-channel full-color-polarization images were taken by the division-of-time polarimeter approach. Then, the CPFA raw data were synthesized with the corresponding CPFA pattern. We quantitatively evaluated with Color Peak Signal-to-Noise Ratio (CPSNR) and the angle error of AoP following related works of~\cite{morimatsu2020monochrome, 9583277}.

Tokyo Tech dataset includes 40 scenes of 1024 $\times$ 768 resolution. 
The evaluations in this work were conducted on our splits: 30 scenes for the training set, two scenes for the validation set, and eight scenes for the testing set.

\subsection{Training details}
\label{sec:training}

We trained the networks with $64 \times 64$ sized patches cropped from the full size of the training image, while we tested with the full size of the test image. For the training batch generation, first, we randomly sampled six different images. Then, four image patches were cropped from each sampled image. Therefore, one training batch for updating the network weight consists of 24 image patches. We used Adam optimizer~\cite{kingma2014adam} with a fixed learning rate of $10^{-4}$ through the training. Each model was trained with 200,000 iterations. Our code is available at our project page~(see fotnote 1).

For data augmentation, we applied the random rotation of $0^{\circ}$, $90^{\circ}$, $180^{\circ}$ and $270^{\circ}$ on the 12-channel ground-truth full-color-polarization images before CPFA raw data synthesis. Applying clockwise $90^{\circ}$ rotation on the polarization image is equivalent to rotating the camera by clockwise $90^{\circ}$ around its Z-axis, which results in the change of AoP. For this case, every pixel in the AoP image should be subtracted by $90^{\circ}$~\cite{kalra2020deep}. We took this AoP change into account by re-arranging the order of polarization channels of the 12-channel ground-truth full-color-polarization images from $[I_0, I_{45}, I_{90}, I_{135}]$ to $[I_{90}, I_{135}, I_{0}, I_{45}]$. Similar polarization channel re-arrangements were also applied for clockwise $180^{\circ}$ rotation and clockwise $270^{\circ}$ rotation.

\subsection{Ablation study}
\label{sec:ablation}

We evaluated two network architectures and two losses in different color spaces. We compared the proposed TCPDNet with the single-step color-polarization network. We also compared the $L_{\text{CP}}$ loss in the RGB color space and the $L_{\text{CP}}^{\text{YCbCr}}$ loss in the YCbCr color space. Our ablation study was conducted with Tokyo Tech dataset.

From Table~\ref{tab:ablation_morimatsu}, we find that the $L_{\text{CP}}^{\text{YCbCr}}$ loss does not contribute to the improvement of the single-step network, but it greatly improves the performance of TCPDNet. This is expected because the single-step network jointly generates the full-color-polarization images originally, while the proposed TCPDNet generates those per monochrome. Thus, the $L_{\text{CP}}^{\text{YCbCr}}$ loss plays an important role in exploiting the inter-channel correlations in our TCPDNet to boost the performance.

\subsection{Comparison with existing methods}
\label{sec:comparison}

We compared our proposed TCPDNet trained with the loss combination $L_{\text{C}} + 4L_{\text{CP}}^{\text{YCbCr}}$ against other existing methods on Tokyo Tech dataset. We compared with five algorithms; bilinear interpolation, EARI~\cite{morimatsu2020monochrome}, IGRI2~\cite{9583277}, CPDNet~\cite{wen2019convolutional} (original), and CPDNet~\cite{wen2019convolutional} (re-trained).
The weight of CPDNet (original) is provided by the authors of~\cite{wen2019convolutional}, while we re-trained CPDNet (re-trained) with Tokyo Tech dataset.  For the learning-based methods, each model was trained five times and averaged metric values were evaluated. 

Table~\ref{tab:sota_morimatsu} shows the quantitative comparisons with Tokyo Tech dataset, where a higher CPSNR is better and a lower angle error is better. We evaluated four color-polarization images ($I_0$, $I_{45}$, $I_{90}$, and $I_{135}$), three Stokes parameters ($S_0$, $S_1$, and $S_2$), DoP, and AoP. From Table~\ref{tab:sota_morimatsu}, we can find that the proposed TCPDNet clearly outperforms the other existing methods by a large margin.

We conducted the qualitative comparison on $S_0$ and the visualizations of DoP and AoP. Figure~\ref{fig:comparison} visualizes the results of different methods for a scene from Tokyo Tech dataset, where we visualize AoP-DoP with the same manner as~\cite{morimatsu2020monochrome, 9583277}. Our proposed TCPDNet can produce better results with clearer edges and  fewer artifacts. On the other hand, we can observe obvious color artifacts in $S_0$ estimated by existing methods, especially on the character "S". Regarding the results of the AoP-DoP visualization, EARI and IGRI2 hardly preserve the edge information. The re-trained CPDNet generally provides better results but not as close to the ground truth as TCPDNet.

\section{Conclusion}
\label{sec:conclusion}

In this work, we have proposed a two-step color-polarization demosaicking network, referred as TCPDNet. This network comprises of two sub-networks: color demosaicking network and polarization demosaicking network. We have also introduced the reconstruction loss in the YCbCr color space for the improvement of TCPDNet. Our proposed TCPDNet quantitatively and qualitatively outperforms existing methods by a significant margin. Comparing to existing methods, our TCPDNet is the best in preserving the edge information with the least color artifacts.

\bibliographystyle{IEEEbib}
\bibliography{refs}

\end{document}